\documentclass[superscriptaddress,nofootinbib,nobibnotes,amsmath,amssymb,aip,preprint]{revtex4-2}
\usepackage{graphicx}
\usepackage{dcolumn}
\usepackage{bm}
\usepackage[T1]{fontenc}
\usepackage{mathptmx}
\usepackage{color}
\usepackage{natbib}
\usepackage{mathtools}
\usepackage{physics}

\begin{document}

\preprint{APS/123-QED}

\title{Giant acoustically-induced synthetic Hall voltages in graphene}

\author{Pai Zhao}
\affiliation{Center for Hybrid Nanostructures, Universit\"at Hamburg, Luruper Chaussee 149, 22761 Hamburg, Germany}
\author{Chithra H. Sharma}
\affiliation{Center for Hybrid Nanostructures, Universit\"at Hamburg, Luruper Chaussee 149, 22761 Hamburg, Germany}
\author{Renrong Liang}
\affiliation{School of Integrated Circuits, Tsinghua University, 100084 Beijing, China}
\author{\textcolor{black}{Christian Glasenapp}}
\affiliation{Center for Hybrid Nanostructures, Universit\"at Hamburg, Luruper Chaussee 149, 22761 Hamburg, Germany}
\author{Lev Mourokh$^{\mathrm{a})}$}
\affiliation{Department of Physics, Queens College of the City University of New York, Flushing, NY 11367, USA}
\author{Vadim M. Kovalev$^{\mathrm{a})}$}
\affiliation{A.V. Rzhanov Institute of Semiconductor Physics, Siberian Branch of Russian Academy of Sciences, Novosibirsk 630090, Russia}
\affiliation{Novosibirsk State Technical University, Novosibirsk 630073, Russia}
\author{Patrick Huber}
\affiliation{Center for Hybrid Nanostructures, Universit\"at Hamburg, Luruper Chaussee 149, 22761 Hamburg, Germany}
\affiliation{Institute of Materials and X-Ray Physics, Hamburg University of Technology, 21073 Hamburg, Germany}
\author{Marta Prada}
\affiliation{I. Institute for Theoretical Physics, Universit\"at  Hamburg HARBOR, Geb. 610 Luruper Chaussee 149, 22761 Hamburg, Germany}
\author{Lars Tiemann\footnote{\footnotesize{correspondences: Lars Tiemann (latieman$@$physnet.uni-hamburg.de) for general correspondence,  Renrong Liang (liangrr$@$mail.tsinghua.edu.cn) for hybrid wafer technology, Lev Mourokh (Lev.Murokh$@$qc.cuny.edu) and Vadim Kovalev (vmk111$@$yandex.ru) for theory.}}}
\affiliation{Center for Hybrid Nanostructures, Universit\"at Hamburg, Luruper Chaussee 149, 22761 Hamburg, Germany}
\author{Robert H. Blick}
\affiliation{Center for Hybrid Nanostructures, Universit\"at Hamburg, Luruper Chaussee 149, 22761 Hamburg, Germany}

\date{\today}
\newpage

\begin{abstract}

Any departure from graphene's flatness leads to the emergence of artificial gauge fields that act on the motion of the Dirac fermions through an associated pseudomagnetic field. Here, we demonstrate the tunability of strong gauge fields in non-local experiments using a large planar graphene sheet that conforms to the deformation of a piezoelectric layer by a surface acoustic wave. The acoustic wave induces a longitudinal and a giant synthetic Hall voltage in the absence of external magnetic fields. The superposition of a synthetic Hall potential and a conventional Hall voltage can annihilate the sample's \textcolor{black}{transverse} potential at large external magnetic fields. Surface acoustic waves thus provide a promising and facile avenue for the exploit of gauge fields in large planar graphene systems.
\end{abstract}

\vfill

\maketitle

\newpage
Phonons and their interaction with electrons have a long and rich history in condensed matter physics. Phonons generally exist as random and omnidirectional propagating lattice vibrations that are powered by the finite temperatures in the solid. 
Surface acoustic waves, henceforth abbreviated as SAW, are very strong directed vibrational modes that propagate predominately along the surface of a piezoelectric crystal. SAW can be launched in a piezoelectric substrate by an interdigitated transducer (IDT)~\cite{1965RWhite_APL}, i.e., two interlocking electrodes connected to a frequency generator, which locally 'nudges' the piezoelectric substrate to perform a periodic mechanical contraction and expansion. The resulting directed propagation of a mechanical deformation wave along the surface of a substrate has become a heavily-exploited technique in sensing applications~\cite{2019PDelsing_JPDAP, 2018SOkuda_ACS_Sensors} and in basic research of 2D electron systems~\cite{1986AWixforth_PRL, 1989AWixforth_PRB, 1990AEfros_PRL, 1990RWillet_PRL, 1992AEsslinger_SSC, 1993VFalko_PRB, 1994AEsslinger_Surf_Sci, 1995DRitchie_JPCM, 1995DRitchie_PRB, 1996SSimon_PRB, 1998AWixforth_APL} in semiconductor heterostructures.

The interaction between SAW and the quasi-relativistic carriers in graphene marries classical and relativistic physics and should significantly revise the acoustic transport of 2D carrier systems. Theory predicts new rich physics, including a crossover from a Schr\"odinger-like to a Dirac-like behavior that relates to the acoustic wave vector\cite{2010PThalmeier_PRB}. When graphene is placed on a substrate with passing SAW, the propagating wave drags along carriers in the graphene sheet. 
Pioneering works have studied the flow of such acoustically induced currents and the buildup of longitudinal voltages under various experimental conditions (temperature, carrier concentration, for different piezoelectric substrates and wiring schemes)~\cite{2012VMiseikis_APL, 2013GNash_APL, 2016GNash_Nano_Res, 2018PSantos_JPDAP, 2018JPollanen_JAP}. Recently, acoustic transport under high magnetic fields and 4.2 Kelvin studied the regime of Landau quantization\cite{2020PZhao_APL} as many years before in GaAs heterostructures\cite{1986AWixforth_PRL,1990RWillet_PRL}.

Here, we explore the limits of very high SAW excitation \textcolor{black}{powers} in the regime of Dirac physics in a large planar graphene device at cryogenic temperatures. We observe the build-up of a giant longitudinal voltage and a synthetic \textcolor{black}{transverse Hall voltage} in the absence of magnetic fields. The acoustically-induced synthetic Hall voltage can be exploited to compensate and amplify conventional Hall voltages that arise from classical electrodynamics when carriers are placed in a magnetic field. We argue that SAW-induced strain leads to artificial gauge fields which can account for the generation of the \textcolor{black}{transverse} voltage component.

\newpage

\begin{figure}[!]
   \includegraphics[width=0.9\textwidth]{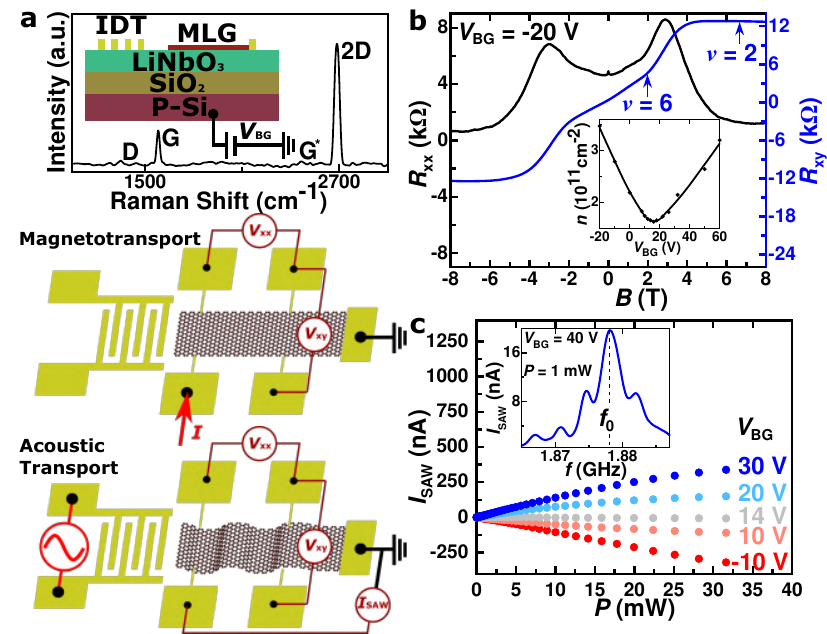}  
   \caption{\text{Device structure, measurement configuration and characterization.} (a) The top main panel shows a Raman spectrum of our CVD monolayer graphene (MLG). The inset is a schematic diagram of the MLG on top of a hybrid LNOI substrate (900 nm of LiNbO$_3$ on 2 $\mu$m SiO$_2$ Insulator). The carrier concentration is tuned by a voltage $V_\mathrm{BG}$ applied to the $p$-doped substrate. MLG was patterned into a Hall bar of 50 $\mu$m width and 400 $\mu$m total length (50 $\mu$m contact distance). The bottom panels illustrate two measurement configurations. In conventional magnetotransport, a constant current, $I$, is passed through the Hall bar. In acoustic transport, a microwave frequency is applied to the IDT (finger width and distance are both 400 nm) and the resulting acoustic current, $I_\mathrm{SAW}$, is measured in short-circuited configuration. In both cases, longitudinal and Hall voltages are measured at the same contacts. (b) Exemplary magneto-transport measurement for $V_\mathrm{BG}$ = -20 V. The inset shows the carrier density vs $V_\mathrm{BG}$. (c) $I_\mathrm{SAW}$ as a function of frequency, power and $V_\mathrm{BG}$ for the fundamental resonance of the IDT. A measurement for the first harmonic is available in the Supplementary Material. \textcolor{black}{Data were obtained at 4.2 K.}}
  \label{fig1}
\end{figure}

Figure \ref{fig1} illustrates our sample layout and summarizes the characterization of the graphene layer. A high piezoelectricity of the substrate is key to strongly actuate the graphene by a passing SAW. An optimized hybrid substrate consisting of 900 nm of LiNbO$_3$ that is separated by 2 $\mu$m SiO$_2$ from a $p$-doped silicon wafer was developed. The $p$-doped Si acts as a back gate to control carrier type and concentration in the graphene via the gate voltage, $V_\mathrm{BG}$. LiNbO$_3$ has a 80 times higher piezoelectric constant than GaAs. The thickness of the dielectric layer is adjusted to the acoustic wave length and suppresses any leakage of the SAW into the bulk. CVD Monolayer graphene (MLG) is transferred to the substrate\cite{2017TLyon_APL} and patterned into a Hall bar close to an IDT (sample details in the caption of Fig. \ref{fig1}). From the Raman spectrum in Fig. \ref{fig1}(a), we deduce a ratio of the 2D/G peaks of 3.5, confirming the presence of a single layer. To abate the amounts of contaminants on the surface, it is thermally annealed inside a vacuum probe at 100$^{\circ}$C for two days prior to cool down to 4.2 K in vacuum. \textcolor{black}{In the following, we will present and discuss data collected from this sample. We reproduced and confirmed the experimental observations on a second sample with different design parameters. The corresponding data are available in the Supplementary Material that accompanies this manuscript.}\\

\newpage

~\\

\newpage

Conventional magneto-transport was measured with the IDT grounded using a standard lockin method, schematically illustrated in the lower part of Fig. \ref{fig1}(a) by passing a constant 4 nA (37.3 Hz) alternating current $I$ through a dedicated source contact close to the IDT. A second contact at the opposite end is permanently grounded. In a four-point scheme, we monitor the longitudinal ($V_\mathrm{xx}$) and Hall ($V_\mathrm{xy}$) voltage with lockin amplifiers while sweeping the magnetic field, $B$. The main panel of Fig. \ref{fig1}(b) shows an exemplary measurement of the longitudinal resistance, $R_\mathrm{xx}$ = $V_\mathrm{xx} / I$, and Hall resistance, $R_\mathrm{xy}$ = $V_\mathrm{xy} / I$, under $V_{BG}$ = -20 V, i.e., for a carrier concentration of 3.47 $\times 10^{11}$ cm$^{-2}$ and a mobility $\mu$ = 6385 cm$^{2}$V$^{-1}$s$^{-1}$. A Hall plateau at Landau level filling factor $\nu$ = $\frac{nh}{eB}$ = 2 is observed at $\textsl{B}$ = 6 T. Detailed magneto-transport measurements are available in the Supplementary Material.

We characterized our IDT and the interaction between SAW and graphene as schematically shown in the lower part of Fig. \ref{fig1}(a). A short-circuit configuration is used to enable a steady acoustic current, $I_\mathrm{SAW}$. The IDT is driven by an external frequency generator in amplitude modulation mode (modulation depth 100$\%$) for the detection of the induced acoustic currents and voltages by lockin amplifiers that also provide a reference signal\cite{1992AEsslinger_SSC}. $V_\mathrm{xx}$ and $V_\mathrm{xy}$ are measured at the same contacts as in conventional magneto transport. 

In the following, we will parameterize the SAW strength by the output power of the frequency generator that is given in mW. Firstly, we will concentrate on the acoustic current that is shown in Figure \ref{fig1}(c). \textcolor{black}{An IDT has a specific fundamental resonance $f_0$ that is given by its design parameters. Around $f_0$, the acoustic wave length $\lambda$ matches the finger distance and the piezoelectric material starts to resonate. Weaker harmonic resonances, with multiples of $\lambda$ matching the finger distance can also be observed. Our} IDT resonates at a fundamental frequency of $f_0$ = 1.878 GHz for which $I_\mathrm{SAW}$ is at a maximum [Fig. \ref{fig1}(c) inset] at any given SAW power. $I_\mathrm{SAW}$ is gate-tunable and its sign reflects the carrier type. Measurements in the main panel of Fig. \ref{fig1}(c) demonstrate that $I_\mathrm{SAW}$ is linear and switches sign when we pass the charge neutrality point (CNP) of our graphene device. The acoustic current density \textcolor{black}{$j_\mathrm{SAW}$} can be expressed by \textcolor{black}{$j_\mathrm{SAW}$} $\propto \mu P^* \Gamma \upsilon^{-1}$, with $P^*$ being the effective wave intensity, $\upsilon$ the wave velocity and $\Gamma$ the wave attenuation, all of which are sensitive to the carrier density or, more precisely, the conductivity~\cite{1998AWixforth_APL, 2017CTang_JAP}. The sensitivity arises from the interaction of electric fields in the travelling mechanical surface deformation of the LiNbO$_3$ and mobile carriers in the graphene. Our IDT can also operate with the (first) harmonic resonance of $f_1$ = 2.872 GHz that results in weaker acoustic currents but shows a similar dependence on the SAW power (additional data in the Supplementary Material). 

\newpage

\begin{figure}[!]
    \includegraphics[width=0.4\textwidth]{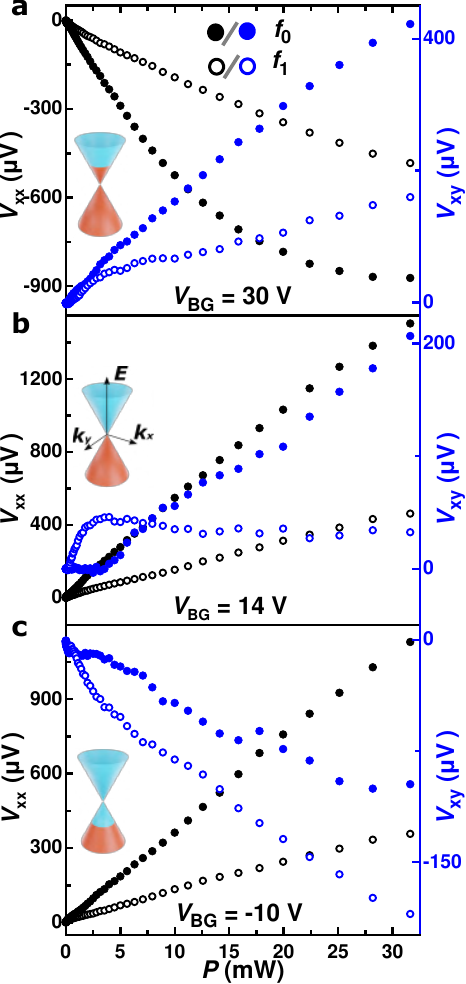}
   \caption{\text{Acoustically induced voltages as a function of power on the IDT.} All measurements are carried out at $B$ = 0 T using the "acoustic transport" configuration of Fig. \ref{fig1}(a). Black always represents longitudinal voltages and blue always \textcolor{black}{transverse} voltages. Closed circles were measured with the fundamental resonance and open circles with the first harmonic of the IDT. From top to bottom, are measurements in the electron regime (a), at CNP (b), and in the hole regime (c) of the MLG. \textcolor{black}{Data were obtained at 4.2 K.}}
  \label{fig2}
\end{figure}~\\ 

\newpage

\begin{figure}[!]
    \includegraphics[width=0.7\textwidth]{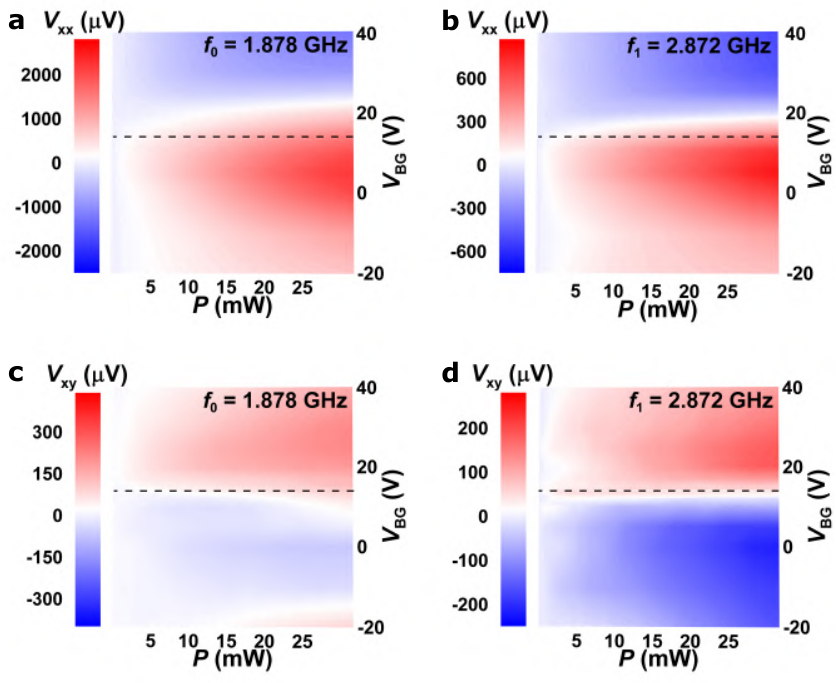}
   \caption{\text{Acoustically induced voltages measured for different gate voltages and IDT powers.} All measurement are performed without external magnetic field at 4.2 Kelvin. (a)/(b) longitudinal voltage for the fundamental and first resonance. (c)/(d) \textcolor{black}{transverse} voltage for the fundamental and first resonance. In all plots, the dashed line marks $V_\mathrm{CNP}$.}
  \label{fig3}
\end{figure}

A directed net flow of carriers that are subject to scattering will result in a potential difference. Consequently, longitudinal voltages were detected in response to the flow of acoustic currents in graphene\cite{2018JPollanen_JAP,2015HKrenner_Nature_Comm}. \textcolor{black}{With increasing IDT power,} we observe that the flow of an acoustic current not only induces the expected longitudinal voltage, $V_{xx}$, but that also a \textcolor{black}{transverse} voltage component, $V_{xy}$, appears. Figure \ref{fig2} shows the development of $V_{xx}$ (black) and $V_{xy}$ (blue) with SAW power for the fundamental and first harmonic resonance of the IDT and for three different carrier concentrations at $B$ = 0 T. The sign of $V_{xx}$ reflects the dominating carrier type in the graphene layer. It is negative for both IDT resonances when the Fermi energy resides in the conduction band and positive when the Fermi energy lies in the valence band. At the CNP, where  $I_\mathrm{SAW}$ becomes vanishingly small, we still observe a large positive $V_{xx}$ of the order of 1 mV. \textcolor{black}{The CNP in our sample is not well-defined and it exhibits a broad transition region [Supplementary Material, Fig. S1a] where electrons and holes co-exist. With 14 V applied to the back gate, holes in the valence band appear to dominate the transport.} The relatively large longitudinal voltage component at the CNP is due to the low conductivity of the graphene layer that enables a large phonon pressure, or a small attenuation of the propagating wave, respectivly. The existence of the \textcolor{black}{transverse} voltage is non-intuitive. It is positive in the electron regime, becomes smaller around the CNP and then changes sign in the hole regime. The more detailed sampling in Fig. \ref{fig3} illustrates that the IDT power is the governing force for all carrier regimes.
 
\begin{figure}[!]
    \includegraphics[width=0.9\textwidth]{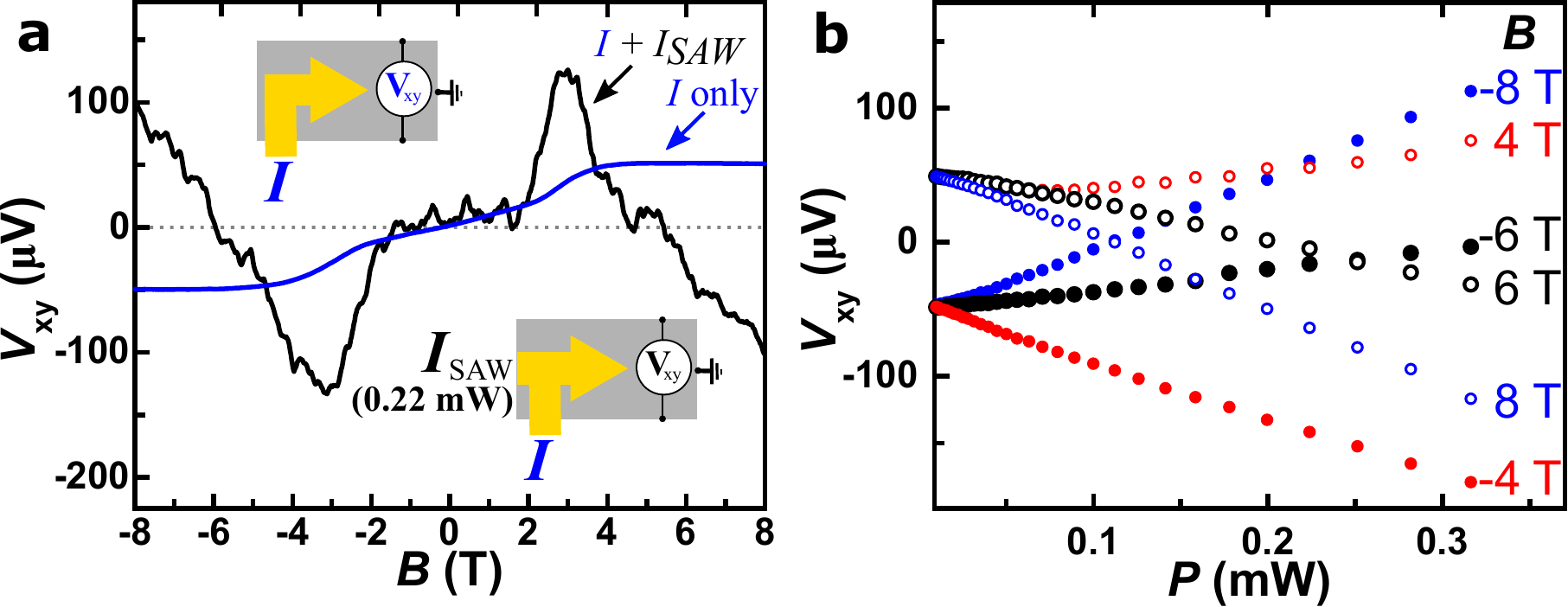}
   \caption{ \textcolor{black}{(a) Ordinary Hall voltage due to a conventional current of $I=4$ nA (upper inset and blue solid line) and transverse voltage measured for a superposition of $I$ and an acoustic current $I_\mathrm{SAW}$ due to a SAW power of 0.22 mW (lower inset and black solid line). The dotted gray line is a guide to the eye indicating $V_\mathrm{xy}$=0. (b) Transverse voltage measured for a superposition of $I$ and $I_\mathrm{SAW}$ at six different magnetic fields and for a varying IDT power. At $\pm$ 6 T, a power of 0.2-0.25 mW is required to fully compensate the transverse voltage to zero, for example. Increasing the IDT power further will again result in a finite net transverse voltage. Data were obtained at 4.2 K with $V_{BG} =$ -20 V and for an $I_\mathrm{SAW}$ generated at $f_0$.}}
  \label{fig4}
\end{figure}

Prior to discussing the nature of the \textcolor{black}{transverse} voltage, we complement our study with an experiment that truly marries classical, quantum and Dirac physics.
At large perpendicular magnetic fields, our device exhibits the quantum Hall effect, with a Hall plateau in the vicinity of 6 T for $V_\mathrm{BG}$ = -20 V [Fig. \ref{fig1}(b)]. We wire our sample again in the conventional magneto transport configuration and impose an alternating current of $I$ = 4 nA. At the same time, we now also fire the IDT at its fundamental resonance that will drive an $I_\mathrm{SAW}$ and acoustically induce \textcolor{black}{transverse} voltage components. Figure \ref{fig4} shows the total $V_{xy}$ measured vs. $B$ (black solid line). The blue line shows the Hall voltage that was measured in the conventional magneto transport configuration (without SAW) for comparison. We note that while $I$ is constant, $I_\mathrm{SAW}$ will depend on the magnetic field\cite{2020PZhao_APL} even for a constant IDT power because the sample conductivity changes when the density of states condenses into Landau levels. By carefully adjusting the IDT power at a constant magnetic field, we are able to completely compensate the conventional Hall voltage by the acoustically induced \textcolor{black}{transverse} voltage component. The IDT power that is needed to fully compensate the Hall voltage is not perfectly symmetric for $-B$ and $+B$ as shown in Fig.\ref{fig4}(b) for six exemplary fields. We suppose this to be arising from sample inhomogeneities and the details of the local charge accumulation.
 
\newpage

~\\ 
\newpage

The build-up of a conventional Hall potential is the trivial consequence of the Lorentz force acting on charged carriers. Here, \textcolor{black}{however, we could observe giant quasi-Hall voltages in \textsl{absence} of an external magnetic field. Below we put forward and discuss an intriguing explanation that is based on \textsl{gauge fields} which arise from the mechanical deformation of the graphene by the SAW.}

\begin{figure}[!]
    \includegraphics[width=0.7\textwidth]{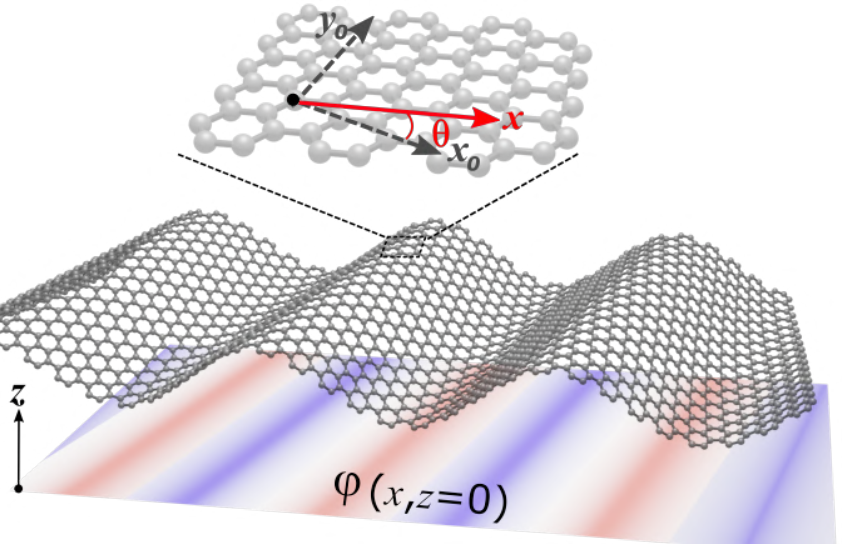}
   \caption{The piezoelectric potential $\varphi(x,z=0)$ in the substrate in contact with the deformed graphene. The direction of propagation is $x || k$ with an arbitrary angle $\theta$ with respect to the crystal axes $x_0$ and $y_0$.}
  \label{fig5}
\end{figure}

Early during the advent of graphene, it was realized that one of many remarkable consequences of its low-energy Dirac description is the emergence of synthetic gauge fields under lattice deformation~\cite{2008Guinea_PRB, 2010MVozmediano_PhysRep, 2017Naumis_ReportPP}. Like regular gauge fields that lead to a magnetic flux, synthetic gauge fields affect the carrier dynamics through an effective pseudomagnetic field $\bm{B} = \bm{\nabla} \times \bm{A}$, with $\bm{A}$ being the  vector potential resulting from a distortion of the graphene lattice\cite{2009FvonOppen_PRB}. Theoretical and experimental work have shown that pseudomagnetic fields can easily exceed tens or even hundreds of Tesla on a nano- and micrometer-scale by strong local deformation~\cite{2010FGuinea_NaturePhysics, 2010NLevy_Science, 2015SZhu_PRL, 2020Sela_PRL}.

Here, we do not rely on sophisticated nanostructuring, however, the \textcolor{black}{transverse} voltage component can be linked to the deformation of the graphene by the SAW and a resulting gauge field~\cite{2015Naumis_JPCM}. We test this hypothesis by considering a defect-free lattice with the crystallographic axes $x_0$ and $y_0$ and a SAW propagation along $x$, as shown in Fig. \ref{fig5}. The SAW, a piezoelectric surface wave in the LiNbO$_3$ substrate travelling with $\upsilon_S$, is accompanied by propagating piezoelectric fields $\varphi(x, z=0) = \varphi_0 \cos(kx-\omega t)$ and a mechanical deformation of the medium that couples/transfers to the graphene layer on its surface at $z$=0. We are interested in the shear component of the mechanical deformation, $u_y(x, z = 0) = u_0 \cos(kx-\omega t)$, associated with the Bleustein-Gulyaev (BG) SAW. The time-dependence of the SAW-induced gauge field has two important consequences: First, in addition to the pseudomagnetic field, there is also the pseudoelectric field, $\bm{E^\mathrm{p}}=-\partial_t\bm{A}$. Both pseudo fields have opposite signs in nonequivalent valleys $K$ and $-K$ of graphene. Second, all the first-order effects associated with the pseudo fields vanish after the time averaging. However, the second-order effects, such as the appearance of the Hall-like field, survive.

We have derived the corresponding synthetic gauge potential in the Supplementary Material. Here, we will only show the components of the resulting pseudoelectric field, $\bm{E^\mathrm{p}}$, and pseudomagnetic fields, $\bm{B^\mathrm{p}}$, in graphene,

\begin{equation}
\bm{E^\mathrm{p}} = A(t)\omega \begin{pmatrix}\sin(3\theta)\\\cos(3\theta)\\0\end{pmatrix}, ~
\bm{B^\mathrm{p}} = A(t) k \begin{pmatrix}0\\0\\\cos(3\theta)\end{pmatrix},
\end{equation}

with $A(t)=\frac{\beta \Delta}{e\upsilon_F}(ku_0)\cos(kx-\omega t)$ being a time-dependent prefactor. Here, $\Delta$ is the hopping amplitude for electrons in the lattice and $\beta$ the Gruneisen parameter for acoustic phonons.

The product of non-zero components (second-order effect) enables a Lorentz-like \textcolor{black}{transverse} force of the form $F_y \propto \langle E^\mathrm{p}_x B^\mathrm{p}_z \rangle$, where $\langle ... \rangle$ indicates time-averaging. The real \textit{piezo}electric field that propagates in the LiNbO$_3$ substrate does not directly contribute to the force on the carriers because it does not change sign in the $-K$ valley. Its product with the pseudomagnetic field vanishes after the summation over the valleys. The product of the pseudoelectric field with the real magnetic field is a first-order effect and also vanishes after time averaging. 

We can now provide a semi-classical estimate of the SAW-induced quasi-Hall voltage on the basis of the balance of forces, as given by

\begin{equation}
  U_H = \mu \langle E^\mathrm{p}_x(t)B^\mathrm{p}_z(t) \rangle d =\frac{\mu d}{4}  \Big(\frac{\beta \Delta}{e\upsilon_F}\Big)^2(k u_0)^2\omega k\sin(6\theta).
\end{equation}

With our experimental parameters of mobility $\mu$ = 0.63 m$^2$V$^{-1}$s$^{-1}$, \textcolor{black}{an assumed} SAW amplitude in the LiNbO$_3$ at low temperatures\cite{2008Shilton_JAP, 2019Islam_APL} $u_0 = 1\cdot 10^{-9}$ m, our fundamental resonance frequency $\omega =2\pi f_\mathrm{0} =  11.7\cdot 10^9$s$^{-1}$, the sample width of $d = 50\cdot 10^{-6}$ m, the Gruneisen parameter\cite{2020Sela_PRL} $\beta \approx 2.5$, the pseudo-vector potential parameter\cite{2020Sela_PRL} $\frac{\Delta}{e\upsilon_F}=2.5 \cdot 10^{-6}$m$\cdot$T and the wave vector $k = 3\cdot 10^6$ m$^{-1}$, we obtain for arbitrary orientations 

\begin{equation}
 U_H \approx 97 \mu \mathrm{V}\cdot \sin(6\theta).
\end{equation}



Our model, based on a balanced (static) force in a perfect crystal, predicts strain-induced synthetic Hall voltages in graphene of the same order that we observed. \textcolor{black}{Below we discuss potential perturbative or spurious parasitic effects in realistic samples which are not captured by this model but which may add to $V_\mathrm{xy}$. A full theoretical analysis of the transverse carrier dynamics still needs to be developed.}

\textcolor{black}{The measured transverse potential represents a local cross section through the width of our CVD graphene Hall bar and encompasses around three or four grains\cite{graphenea}, which is beneficial. Having only a single grain with $\theta$ inadvertently close to 0, 30$^{\circ}$,... would suppress $U_H$. The measured $V_\mathrm{xy}$ thus represents an average. A complete averaging out to zero however would only be possible in the limit of a very high number of grains. Any departure from flatness introduced in
the fabrication process also leads to inhomogeneous gauge field contributions\cite{2010MVozmediano_PhysRep}.}

\textcolor{black}{We attribute nonlinearities in $V_\mathrm{xy}(P)$ around the CNP to the coexistence and fluctuations of electron-hole puddles. These density inhomogeneities smear out and broaden the CNP [Supplementary Material Fig. S1a]. Additional small contributions may arise from spurious reflected attenuated acoustic waves that have bounced off from the edges or the bottom of the substrate. Also, the coupling between substrate and graphene may vary locally.}

\textcolor{black}{Contributions from skew scattering will exist but should be insignificantly small. The mean free path is small compared to the sample width. In contrast to the net deflection by a Lorentz force, skew scattering will occur randomly on defects which cannot result in a large charge accumulation.}

\textcolor{black}{Despite all these possible spurious contributions, the strong deformation by the SAW appears to dominate and we consistently observe strong acoustic voltages in phase to the currents and only at the IDT resonances. We reproduced these findings on a second sample with different design parameters [Data in the Supplementary Material] so that the most probable scenario for the observation of synthetic Hall potentials under a SAW is the emergence of gauge fields.}\\

Our study demonstrates that the propagation of a surface acoustic wave in a highly piezoelectric substrate induces large longitudinal and synthetic Hall potentials in a graphene layer that conforms to its surface. The magnitude and sign of the acoustic voltages can be controlled by the carrier concentration and carrier type in the graphene as well as the intensity of the SAW in the supporting substrate. The emergence of the synthetic Hall potential is consistent with gauge fields, induced by the mechanical deformation of the propagating SAW. The observed gauge fields are extremely strong, empowering us to acoustically compensate the conventional Hall voltage that builds up under strong magnetic fields. The high accessibility of these experiments may highlight a new arena for easy and effective control of gauge fields in graphene in fundamental research and applied sciences.\\\\\\

\noindent \textbf{Addendum:}\\

After the publication of our manuscript [P. Zhao \textsl{et al.}, Phys. Rev. Lett. 128, 256601 (2022)], we became aware of related work simultaneously published by P. Bhalla \textsl{et al.} [Phys. Rev. B 105, 125407 (2022)]. Using Kubo's formalism, the authors theoretically studied 'acoustogalvanic effects' in two-dimensional Dirac materials that arise from sound-induced pseudo-gauge fields. Despite the differences in our two models, their predicted acoustoelectric responses resemble our experimental observations.\\

\newpage

\noindent \textbf{Acknowledgement:}\\
R.H.B. acknowledges support by the Deutsche Forschungsgemeinschaft under grant BL-487/14-1. P.Z. acknowledges financial support by the Zentrum für Hochleistungsmaterialien ZHM, Hamburg, Germany. R.R.L. acknowledges the funding of the National Natural Science Foundation of China (No. 92064002). \textcolor{black}{V.M.K. acknowledges support by the Foundation for the Advancement of Theoretical Physics and Mathematics 'BASIS'}. C. H. S. acknowledges financial support from the Alexander von Humboldt foundation. P. H. acknowledges support by the Deutsche Forschungsgemeinschaft under grant no. 192346071, SFB 986 “Tailor-Made Multi-Scale Materials Systems”, as well as by the Center for Integrated Multiscale Materials Systems CIMMS, funded by Hamburg science authority. All measurements within the scope of this work were performed with \textsl{nanomeas}.\\\\

\newpage

\bibliographystyle{plainnat}

\newpage

\begin{center}
	\begin{LARGE}{\textbf{Supplementary Material}}\end{LARGE}\\
	for\\
	Giant acoustically-induced synthetic Hall voltages in graphene\\
	by\\
	Pai Zhao \textsl{et al.}
\end{center}

\section[1.]{SAMPLE AND SETUP}

\subsection[1.1.]{Samples}

We have prepared and measured two samples. Sample SA, whose data are shown the main text, is a CVD graphene Hall bar [length 400 $\mu$m, width 30 $\mu$m, contact distance 50 $\mu$m] on a 900 nm LiNbO$_3$ on 2$\mu$m SiO$_2$ insulator (to enable gating via a backgate) with a Y-cut of the substrate. We prepared an IDT with a finger width and distance of both 400 nm, resulting in a fundamental resonance of $f_0$ = 1.878 GHz. This emperical resonance matches the expected resonance based on the speed of sound in LiNbO$_3$. A schematic diagram of SA is show in Fig.1 in the main text. The control sample SB is a CVD graphene Hall bar [length 1 mm, width 400 $\mu$m, contact distance 200 $\mu$m and 600 $\mu$m] on \textbf{bulk} 450 $\mu$m LiNbO$_3$, i.e., no insulator, with a 128°-Y cut. Here, the IDT finger width and distance are both 2 $\mu$m, resulting in a lower fundamental resonance $f_0$ = 481.5 MHz, which also matches our estimate. These two very different wafers/IDTs were selected to exclude possible artifacts arising from the SAW  design and frequency range. The larger $d$ partially compensates the smaller $\omega$ and $k$ [equ. (9)]. The monolayer CVD graphene of SA (SB) has an intrinsic carrier concentration of $p=$2.11$\times$10$^{11}$ (1.73$\times$10$^{12}$) cm$^{-2}$ and a mobility of 6274 (2623) cm$^2$V$^{-1}$s$^{-1}$. Sample SB has one additional pairs of Hall contacts but no back gate (i.e., all measurements for SB below are shown for its intrinsic carrier concentration).

\subsection[1.2.]{Standard Magneto-Transport}
Magneto transport measurements were always performed using a four-point method by applying a constant (alternating) current of 4 nA and 37.3 Hz between a source and a drain contact. Longitudinal and Hall voltages were measured with Stanford lockin amplifiers at separate contacts.

\subsection[1.3.]{Acoustic Measurements}

For all acoustic measurements (except for the \textit{compensation} measurements) no external constant alternating current is applied to the graphene Hall bar and the magnetic field is zero at all times. The IDT is connected to a frequency generator that operates in AM mode (i.e., amplitude modulation, modulation depth 100$\%$), receiving a 37.3 Hz TTL reference signal from a dedicated lockin amplifier. In all measurements, we naturally ensure that the lockin out-of-phase components (y-components) are zero.\\

We can use both \textit{open-circuit} and \textit{short-circuited} configurations for the electrical measurements of the graphene Hall bar under SAW excitations. To enable a steady flow of an acoustic current, a \textit{short-circuited} configuration is needed that closes the electrical circuitry. In the \textit{short-circuited}, a lockin measures the induced acoustic current at a contact close to the IDT while a contact at the opposite end of the Hall bar is grounded to close the electrical circuit (sample SA). The longitudinal and transverse voltage are always measured at the same contacts as in conventional magneto transport. In the \textit{open-circuit} configuration with only one or no contact grounded, charges can still accumulate and generate a potential difference (voltage) but a steady acoustic current cannot flow.\\

\noindent \textit{Compensation} measurements:\\
\indent In the \textit{compensation} measurements, an external constant alternating current is superimposed onto an acoustic current. The constant (alternating) current of $I$ = 4 nA (37.3 Hz) is passed through the Hall bar while simultaneously the IDT is triggered by a frequency generator in AM modulation mode (via a reference signal from the lockin) which induces an acoustic current $I_\mathrm{SAW}$. There exists no phase difference between the constant low frequency ac of 37.3 Hz and the (low frequency) 37.3 Hz envelope of the modulated radio frequency. In this experiment, we use the same contacts as in in the individual experiments.

\newpage

\section[2.]{Additional experimental data from sample SA}

\subsection[2.1.]{Standard Magneto-Transport (Sample SA)}

A maximum in the longitudinal resistance measured for $B = 0$ T as a function of the back gate voltage, $V_\mathrm{BG}$, indicates the charge neutrality point (CNP), as shown in Fig. S1(a). Figures S1(b) and S1(c) show the longitudinal and Hall resistances in $B$ - $V_\mathrm{BG}$ space. A clear Landau level filling factor $\nu$ = 2 is observed as the formation of a Hall plateau and in the development of minina in the Shubnikov de Haas oscillations. Figure S1(d) analyzes the slope of the Hall resistance, indicating three distinguished regimes reflecting the transition from the hole regime (light red) to a two carrier regime around the CNP of both holes and electrons and to the electron regime (light blue). The slope of the Hall resistance switches sign at $V_\mathrm{CNP}$, which is consistent with Fig. S1(a).

\begin{figure}[!ht]
	\includegraphics[width=0.8\textwidth, angle=0]{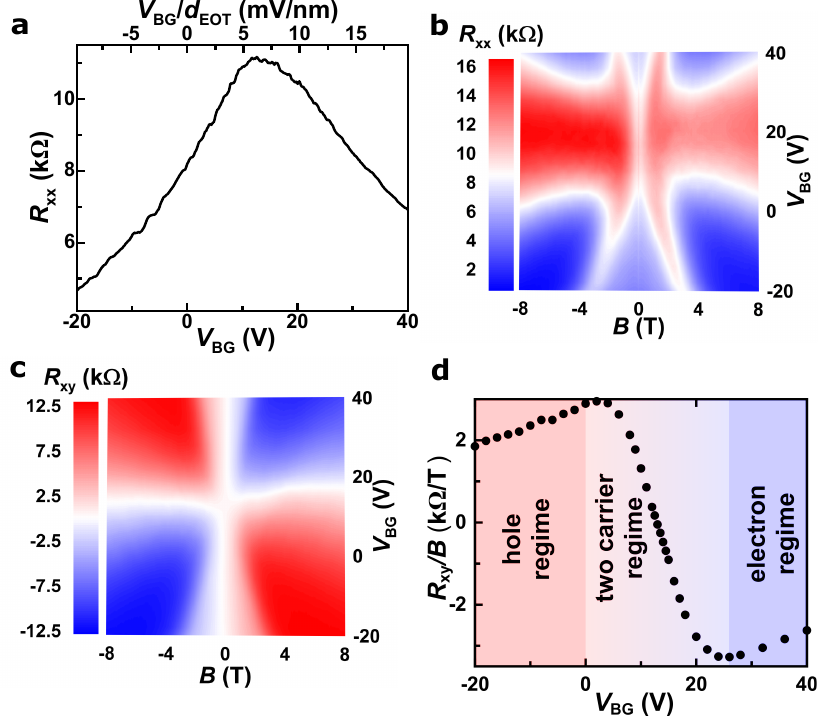}
	\caption{Magnetotransport measured with a constant (alternating) current,  $I$ = 4 nA, at 4.2 K on sample SA. (a) longitudinal resistance measured as a function of back gate voltage from -20 to 40V at $B$ = 0 T. The upper axis shows the applied voltage calibrated to $d_\mathrm{EOT}$ (effective oxide thickness) including the dielectric thickness of 900 nm LiNbO$_3$ and 2 $\mu$m SiO$_2$. (b) and (c) longitudinal and Hall resistances measured as a function of magnetic field and back gate voltage. A clear transition from hole to electron regime is observed. (d) The slope of the Hall resistance signals the regimes of pure electron transport, the CNP region and hole transport.}
	\label{figS1}
\end{figure}


\subsection[2.2.]{Acoustic currents (Sample SA)}

Figure 1(c) in the main text illustrates the acoustic current for the fundamental resonance $f_0$ of the IDT (sample SA). In Fig. \ref{figS2} below, we show additional data from the same sample. $I_\mathrm{SAW}$ was determined as a function of $V_\mathrm{BG}$ for a constant power of 3.16 mW (5 dBm). $R_\mathrm{xy}/B$ was determined by a conventional electric field driven current (conventional magneto transport) and shows nearly perfect agreement with the transition from electron to hole regime. The small difference in the zero-crossings of $I_\mathrm{SAW}$ and $R_\mathrm{xy}/B$ of $V_\mathrm{CNP}$ (0.68 mV/nm) might be induced by strain from the passing SAW that redistributes electron-hole puddles close to CNP.

\begin{figure}[!hb]
	\includegraphics[width=0.6\textwidth, angle=0]{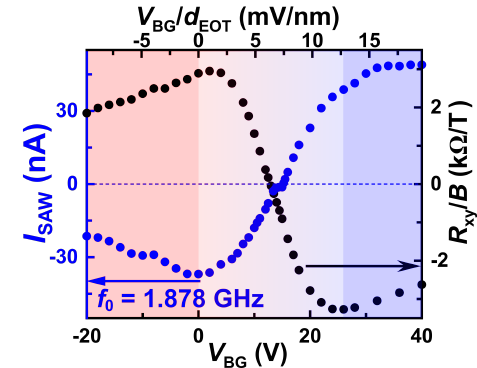}
	\caption{Acoustic current $I_\mathrm{SAW}$ for $P$ = 3.16 mW and various gate voltages (blue dots) for the fundamental resonance $f_0$ at 4.2 K (sample SA). Magnitude and sign of $I_\mathrm{SAW}$ reflect the transition from electron to hole regime. The slope of Hall resistance (black dots, adapted from magnetotransport), shows the same transition voltage.}
	\label{figS2}
\end{figure}

\newpage

\begin{figure}[!hb]
	\includegraphics[width=0.6\textwidth, angle=0]{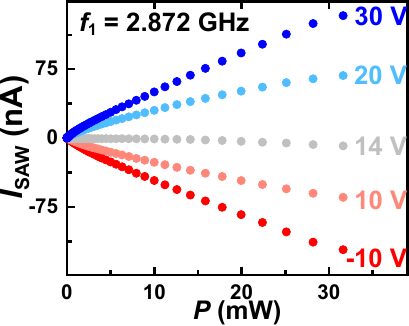}
	\caption{Acoustic current measured at the first harmonic, $f_1$ = 2.872 GHz for various powers and selected gate voltages at 4.2 K (sample SA).}
	\label{figS3}
\end{figure}

Figure \ref{figS3} shows complementary measurements for the first harmonic $f_1$ = 2.872 GHz as a function of the power to the IDT for selected gate voltages (sample SA). We observe the same linear dependence as in Fig. 1(c) [main text] but with a reduced magnitude. 

\newpage

\subsection[2.3.]{SAW-induced synthetic Hall voltage (SAMPLE SA)}

\begin{figure}[!ht]
	\includegraphics[width=0.45\textwidth, angle=0]{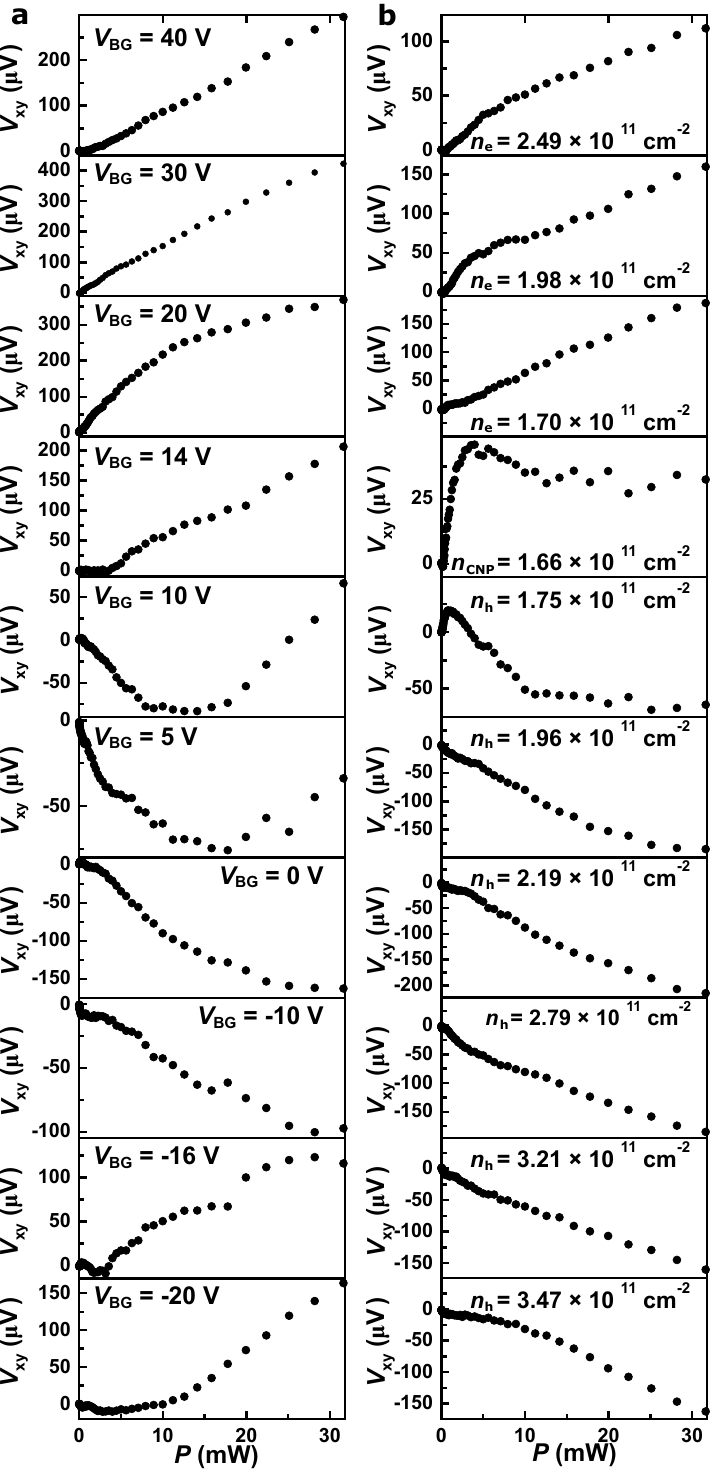}
	\caption{Development of SAW-induced synthetic Hall voltage with carrier type, carrier concentrations and intensity of SAW for (a) the fundamental resonance $f_0$ and (b) the first harmonic $f_1$. Each voltage in (a) yields the carrier concentration shown in (b), i.e., carrier concentrations do not differ between $f_0$ and $f_1$. All data from sample SA and at 4.2 K.}
	\label{figS4}
\end{figure}

Figure \ref{figS4} shows additional representative individual plots of SAW-induced transverse voltage at different carrier concentrations (sample SA). Subfigures (a) [left column] are data for the fundamental resonance $f_0$ and subfigures (b) [right column] for the first harmonic $f_1$.

~\\

A strong nonmonotonic behavior in $V_\mathrm{xy}$ occurs around the CNP. Our CVD graphene exhibits a broad CNP (Fig S1a), where electron and hole puddles coexist. In this regime, the electron-hole landscape between the Hall probes is sensitive to the SAW power and $V_\mathrm{xy}$ fluctuates with increasing $P$.

\subsection[2.4.]{Compensation measurement out-of-phase components (Sample SA)}

\begin{figure}[!ht]
	\includegraphics[width=0.5\textwidth, angle=0]{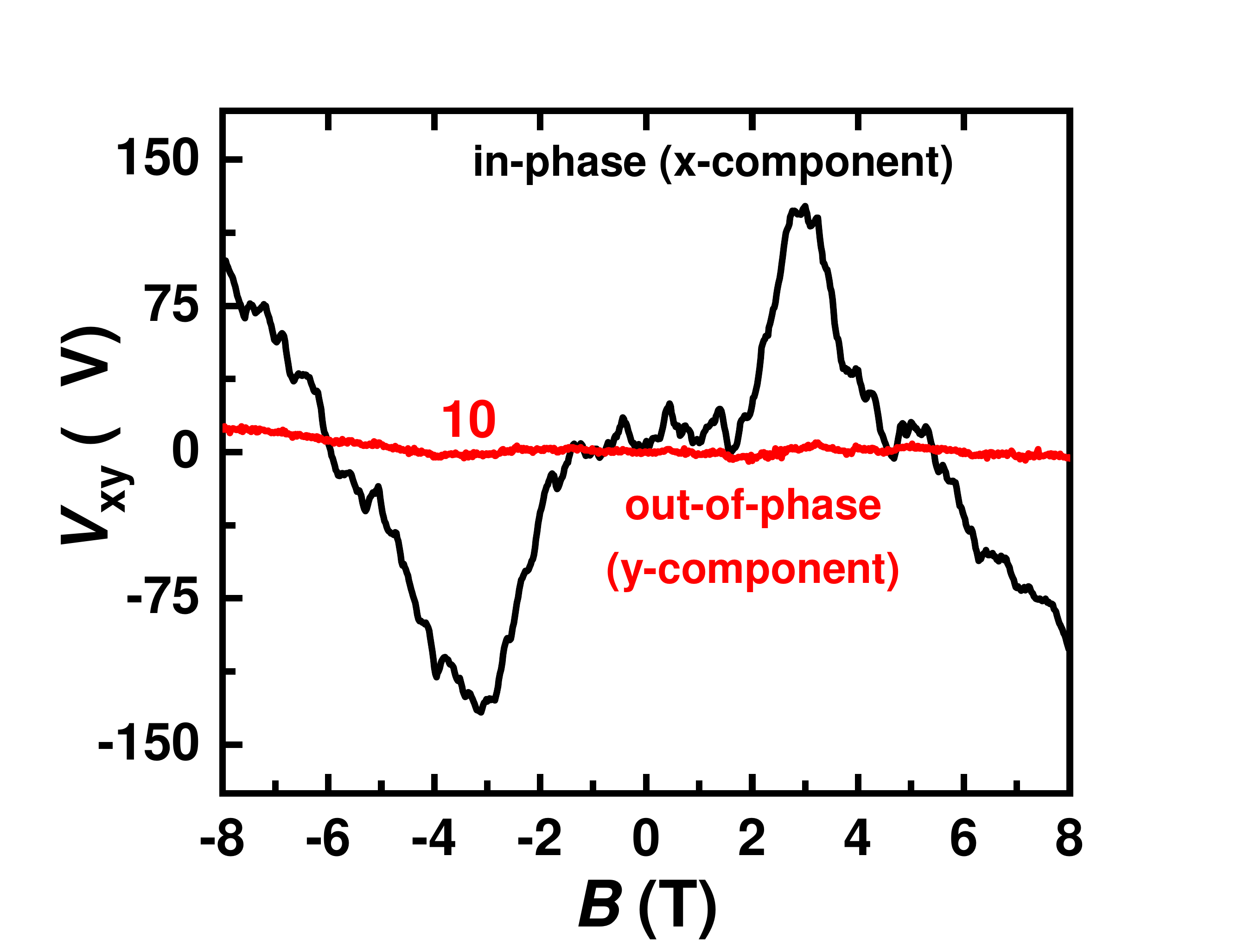}
	\caption{The solid black line is identical to Fig.4 a in the main text and represents the in-phase (x) component of the lockin amplifier for $f_0$ with a power of 0.22 mW. The solid red line shows the corresponding out-of phase (y) component multiplied by a factor of 10. Data from sample SA and at 4.2 K.}
	\label{figS5}
\end{figure}

The high frequency signal from the frequency generator that triggers the IDT is modulated with a 37.3 Hz signal from a lockin amplifier. Both signals are phase-locked and there exists no phase-difference between the conventional 37.3 Hz ac current $I$ and the envelope of the modulated signal that triggers the $I_\mathrm{SAW}$. Figure \ref{figS5} further demonstrates that our measurement signals that are detected by the lockin amplifiers do not exhibit an out-of-phase (y) component. The solid black line in Figure \ref{figS5} is identical to that in Fig. 4a of the main text. It shows the in-phase (x) component of the lockin amplifier. The solid red line is the out-of-phase (y) component  that was multiplied by a factor of 10 in order to see the negligible variations from zero. The compensation of $V_\mathrm{xy}$ to zero is thus not the result of a continuously changing phase in the measurement signal but due to the superposition of the two in-phase components of the conventional Hall voltage and the synthetic transverse voltage. 

\newpage

\section[3.]{Additional experimental data from the control sample SB}

Here, we present the most relevant data obtained from our control sample SB that qualitatively confirms the observations made in sample SA. We note that for the SAW experiments, we could employ smaller IDT powers because the semi-rigid coaxial wire that connects the frequency generator to the IDT has a considerably higher transmission in this frequency range. This was determined by a characterization measurement with a Vector Network Analyzer (VNA). Hence, we obtain similar effects with effectively lower absolute IDT powers. All data are obtained at 4.2 Kelvin and with the intrinsic (hole) concentration of 1.73 $\times$ 10$^{12}$ cm$^2$/Vs.\\

\begin{figure}[!ht]
	\includegraphics[width=0.5\textwidth, angle=0]{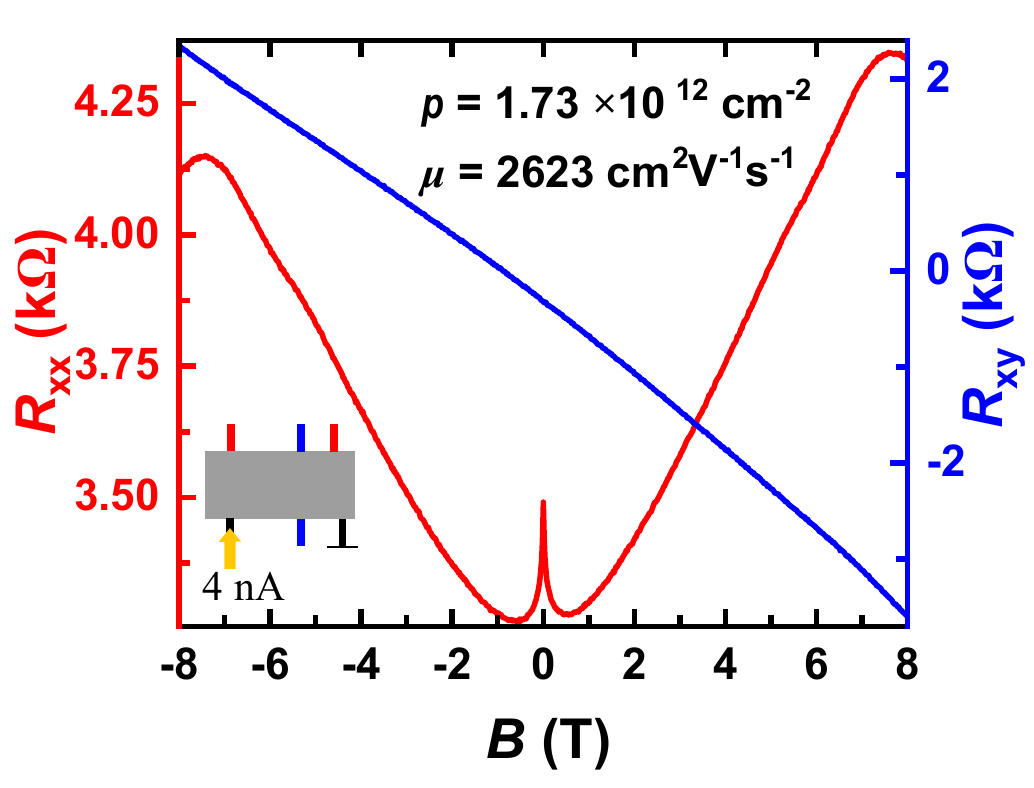}
	\caption{Conventional magneto-transport Data from sample SB with a current of 4 nA. The inset shows a color-coded contact/measurement configuration.}
	\label{figS6}
\end{figure}

Figure \ref{figS6} shows regular magneto-transport performed with a standard lockin method and an low-frequency (37.3 Hz) alternating current of 4 nA.

\begin{figure}[!ht]
	\includegraphics[width=0.95\textwidth, angle=0]{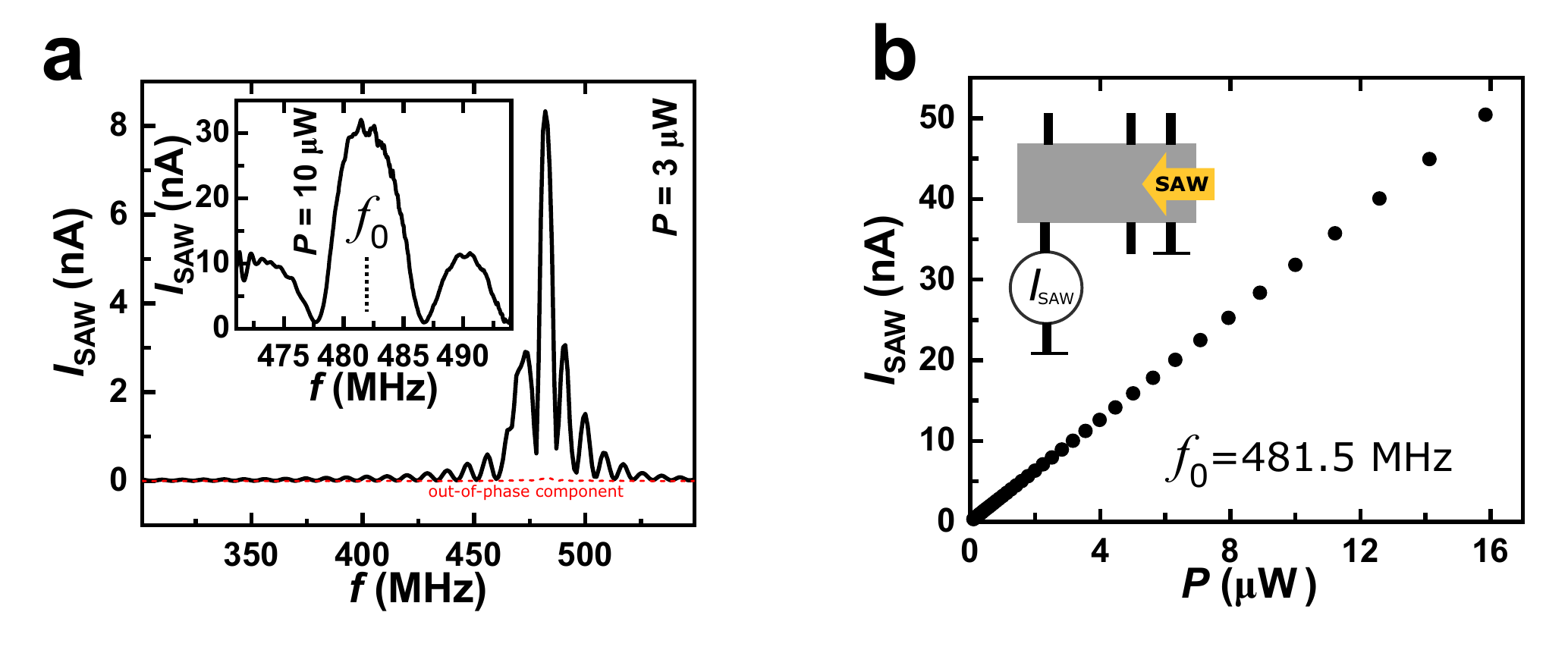}
	\caption{a. (main panel) A low resolution sweep (i.e., large $\Delta f$ steps) of the IDT frequency $f$ and the resulting measured acoustic current (black solid line). The red dashed line is the out-of phase (y) component, which is negligible. The inset shows a high resolution frequency sweep (small $\Delta f$ steps) around the fundamental resonance, $f_0$ = 481.5 MHz. $P$ always represents the power applied to the IDT. b. Acoustic current as function of the IDT power $P$ at $f_0$. The inset shows the contact/measurement configuration used for the data shown in subfigures a and b.}
	\label{figS7}
\end{figure}

The main panel of Figure \ref{figS7}a shows the acoustic current, which was measured as shown in the inset in subfigure b, as a function of frequency $f$ (black solid line) with a low resolution (i.e., larger $\Delta f$ steps) and low IDT power. The red dashed line is the lockin out-of phase (y) component which is negligibly small. The inset in Figure \ref{figS7}a shows a high resolution (small $\Delta f$ steps) measurement around the fundamental resonance $f_0$ = 481.5 MHz of the IDT with larger power. Figure \ref{figS7}b shows the acoustic current versus power applied to the IDT for $f_0$.\\\\

\begin{figure}[!ht]
	\includegraphics[width=0.95\textwidth, angle=0]{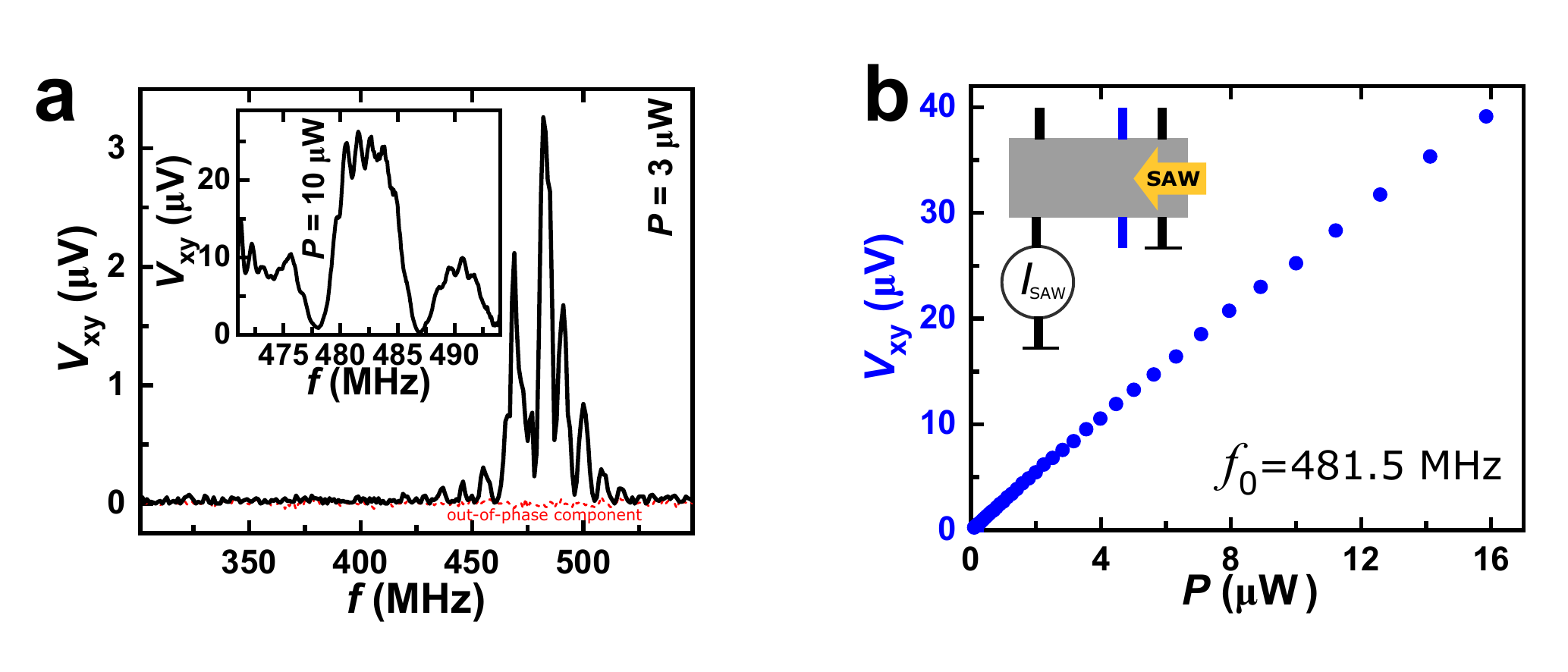}
	\caption{a. (main panel) A low resolution sweep (i.e., large $\Delta f$ steps) of the IDT frequency $f$ and the resulting measured transverse voltage $V_\mathrm{xy}$ (black solid line). The red dashed line is the out-of phase (y) component, which is zero. The inset shows a high resolution frequency sweep (small $\Delta f$ steps) around the fundamental resonance, $f_0$ = 481.5 MHz. $P$ always represents the power applied to the IDT. b. $V_\mathrm{xy}$ as function of the IDT power $P$ at $f_0$. The inset shows the contact/measurement configuration.}
	\label{figS8}
\end{figure}

The main panel of Figure \ref{figS8}a shows the transverse (quasi Hall) voltage, which was measured as shown in the inset in subfigure b, as a function of frequency $f$ (black solid line) with a low resolution (i.e., larger $\Delta f$ steps) and low IDT power. The red dashed line is the lockin out-of phase (y) component which is zero. The inset in Figure \ref{figS8}a is a high resolution (small $\Delta f$ steps) measurement around the same fundamental resonance $f_0$ = 481.5 MHz of the IDT with larger power. Figure \ref{figS8}b shows the transverse voltage versus power applied to the IDT for $f_0$. 

\begin{figure}[!ht]
	\includegraphics[width=0.55\textwidth, angle=0]{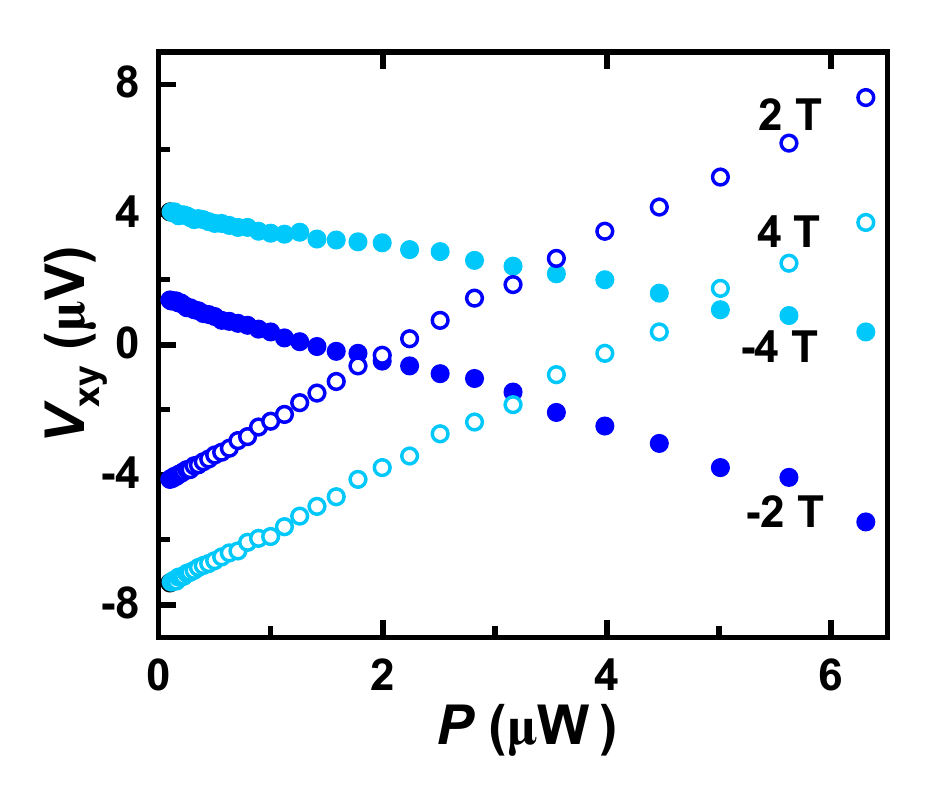}
	\caption{$V_\mathrm{xy}$ when a conventional alternating current of 4 nA (37.4 Hz) is superimposed to the acoustic current that is induced by the SAW. The measurement configuration is a combination of the insets in Figs. \ref{figS6}, \ref{figS7}, and \ref{figS8} as explained in section I.C. At $\pm$2 T and $\pm$4 T, fine-tuning of the IDT power allows to compensate the transverse voltage to zero. }
	\label{figS9}
\end{figure}

Figure \ref{figS9} reproduces the compensation measurement by adding a conventional ac current to an acoustic current. At finite magnetic fields, here we show examples at $\pm$2 T and $\pm$4 T, and for an adjusted IDT power, the transverse voltage vanishes.

\newpage

\section[4.]{Estimation of the Hall voltage}

In this section, we provide the details of the calculations which have led to the estimations of the values for the synthetic Hall voltage induced by SAW. Acoustic waves propagating along the surface of piezoelectric LiNbO$_3$ substrate produce both the mechanical deformation of the media with the displacement vector ${\bf u}({\bf r},t)$ and the piezoelectric field characterized by the scalar potential $\varphi({\bf r},t)$. We examine the geometry of Fig.5, with SAW traveling along $x$-direction (shown in black), while the graphene crystallographic axes $(x_0,y_0)$ are shown in red. The angle between the coordinate systems is $\theta$. In the case of Bleustein-Gulyaev (BG) SAW, the mechanical displacement is perpendicular to the direction of the propagation (shear stress). At the interface between the piezoelectric substrate and the graphene layer the shear displacement is given by 
\begin{gather}\label{Eq1_SM}
u_y(x,t)=u_0\cos(kx-\omega t).
\end{gather}
The piezoelectric potential has the form
\begin{gather}\label{Eq2_SM}
\varphi(x,t)=\varphi_0\cos(kx-\omega t),
\end{gather}
with the amplitudes $u_0$ and $\varphi_0$ coupled by the piezoelectric modulus at the interface. The mechanical deformation due to BG SAW is usually overlooked because its contribution is weaker than that of the piezoelectric potential and, also, it is perpendicular to the direction of acoustic current. However, we will show that it is this component which produces the Hall-like voltage via pseudomagnetic and pseudoelectric fields. 

We assume that there is a good mechanical coupling between the LiNbO$_3$ substrate with the graphene monolayer and the mechanical stress described by equation \eqref{Eq1_SM} is well transferred to the graphene monolayer. It is well known that mechanical deformations induce gauge fields which are analogous to electromagnetic fields that are characterized by the vector potential $\bm{A}$ acting on the 2D graphene electron gas. The corresponding electron Hamiltonian is given by
\begin{gather}\label{Eq3_SM}
H={\bf v}\cdot({\bf p}-e{\bf A})+e\varphi,
\end{gather}
where ${\bf v}=v_F(\sigma_x,\sigma_y)$ is the  velocity operator for graphene electrons. Thus, one can see that the electrons feel a piezoelectric field given as $\bm{\mathcal{E}}=-\bm{\nabla}\varphi$, and artificial pseudo-electromagnetic (pseudoEM) field, having the pseudoelectric $\bm{E^\mathrm{p}}=-\partial_t\bm{A}$ and pseudomagnetic  $\bm{B^\mathrm{p}}=\bm{\nabla}\times\bm{A}$ components. It should be noted that both pseudo field have opposite signs in nonequivalent valleys $K$ and $-K$ of graphene.

In case of an arbitrary orientation of the BG SAW wavevector $\bm{k}$ with respect to the crystallographic system $(x_0,y_0)$, the vector potential $\bm{A}=(A_x,A_y)$ of the pseudoEM field reads
\begin{gather}\label{Eq4_SM}
A_x=\frac{\beta \Delta}{ev_F}\Bigl[(\varepsilon_{xx}-\varepsilon_{yy})\cos(3\theta)-2\varepsilon_{xy}\sin(3\theta)\Bigr],\\\nonumber
A_y=\frac{\beta \Delta}{ev_F}\Bigl[-2\varepsilon_{xy}\cos(3\theta)+(\varepsilon_{xx}-\varepsilon_{yy})\sin(3\theta)\Bigr],
\end{gather}
where $\beta$ is the Gruneisen parameter, $\Delta$ is the hopping integral, and
\begin{gather}\label{Eq5_SM}
\varepsilon_{\alpha\beta}=\frac{1}{2}\left(\partial_\beta u_\alpha+\partial_\alpha u_\beta\right)
\end{gather}
is a stress tensor. In a case of a BG wave, the stress tensor contains only non-diagonal components $\varepsilon_{xy}$ describing the shear deformation. Thus, 
\begin{gather}\label{Eq6_SM}
A_x(x,t)=\left(\frac{\beta \Delta}{ev_F}\right)(ku_0)\sin(kx-\omega t)\sin(3\theta),\\\nonumber
A_y(x,t)=\left(\frac{\beta \Delta}{ev_F}\right)(ku_0)\sin(kx-\omega t)\cos(3\theta).
\end{gather}
The corresponding pseudoelectric and pseudomagnetic components of an artificial EM field are given by
\begin{gather}\label{Eq7_SM}
E_x^\mathrm{p}(x,t)=-\partial_tA_x(x,t)=\omega\left(\frac{\beta \Delta}{ev_F}\right)(ku_0)\cos(kx-\omega t)\sin(3\theta),\\\nonumber
E_y^\mathrm{p}(x,t)=-\partial_tA_y(x,t)=\omega\left(\frac{\beta \Delta}{ev_F}\right)(ku_0)\cos(kx-\omega t)\cos(3\theta),\\\nonumber
B_x^\mathrm{p}=0,\\\nonumber
B_y^\mathrm{p}=0,\\\nonumber
B_z^\mathrm{p}=k\left(\frac{\beta \Delta}{ev_F}\right)(ku_0)\cos(kx-\omega t)\cos(3\theta).
\end{gather}
It is evident that the terms containing these oscillating fields would be averaged out. However, the second order products, such as 
 $\langle\bm{\mathcal{E}}^2\rangle$, $\langle\bm{E}^\mathrm{p2}\rangle$,  $\langle\bm{\mathcal{E}}\cdot\bm{E}^\mathrm{p}\rangle$, $\langle\bm{\mathcal{E}}\cdot\bm{B}^\mathrm{p}\rangle$, $\langle\bm{E}^\mathrm{p}\cdot\bm{B}^\mathrm{p}\rangle$,  $\langle\bm{\mathcal{E}}\times\bm{B}^\mathrm{p}\rangle$,  $\langle\bm{E}^\mathrm{p}\times\bm{B}^\mathrm{p}\rangle$ ($\langle ... \rangle$ means the time averaging), might provide non-vanishing contributions. We are interested in the two latter cases which may induce the Hall-like voltages. The first vector product, $\langle\bm{\mathcal{E}}\times\bm{B}^\mathrm{p}\rangle$, containing the real piezoelectric field and the first order of the pseudoEM field, vanishes after the summation over the valleys because the signs of pseudoEM are different in  $K$ and $-K$ valleys. The non-zero contribution can appear only for broken time reversal symmetry, which is not the case in our system. Both factors in the last term, $\langle\bm{E}^\mathrm{p}\times\bm{B}^\mathrm{p}\rangle$, change the sign in different valleys, so their product remains the same. Full theoretical analysis of the acoustically-induced transverse Hall-like voltage based on the Boltzmann equation with account of electron-impurity scattering processes will be done elsewhere. Here, we provide the estimation of SAW-produced Hall-like voltage on the basis of a force balance.

The Hall-like voltage in $y$-direction of Fig.1 is $U_H=E_Hd$, where $E_H$ is the static Hall electric field and $d$ is the sample width. In the steady state, the force balance yields the relation
\begin{gather}\label{Eq8_SM}
eE_H=e\langle v_x(x,t)B_z^\mathrm{p}(x,t)\rangle,
\end{gather}
where $v_x(x,t)$ is the alternating component of electron velocity due to the pseudoelectric field $E_x^\mathrm{p}(x,t)$ in Eq.\eqref{Eq7_SM}.

For the estimation purposes, we assume that $v_x(x,t)=\mu E_x^\mathrm{p}(x,t)$, where $\mu$ is the electron mobility in graphene. Thus, the Hall-like voltage is given by
\begin{gather}\label{Eq9_SM}
U_H=\mu\langle E_x^\mathrm{p}(x,t)B_z^\mathrm{p}(x,t)\rangle d=\frac{\mu d}{4}\left(\frac{\beta \Delta}{ev_F}\right)^2(ku_0)^2\cdot\omega k \cdot\sin(6\theta).
\end{gather}~\\
This expression is used in the main text to estimate the transverse synthetic voltage. The values for $d$, $\mu$, $\omega$ and thus $k$ are directly given by the sample and IDT design. For $\frac{\Delta}{ev_F}$ and $\beta$ we use values from other published work [\text{Phys. Rev. Lett.} \textbf{124}, 026602 (2020)]. In passing we note that $v_F$ is known to slightly depend on the substrate material. The SAW amplitude $u_0$ is not well-defined and we rely on an estimate based on general studies of our piezoelectric LiNbO$_3$ substrate. $u_0$ is thus the biggest uncertainty in our estimate. Its value increase with SAW power and also changes with the wave length.

$\theta$ is the (unknown) local graphene lattice orientation between the Hall probes. We emphasize that from a sample fabrication standpoint, it is not really possible to control $\theta$. For exfoliated graphene or CVD graphene with only a single grain, equation (9) implies a high probability of having an unfavorable $\theta$ close to 0°, 30°, 60° ..., which would completely suppress $U_H$. Having a small number of grains is thus beneficial as is increases the probability for a non-zero average of $U_H$. A complete averaging out to zero would only be possible for a very large number of grains, however.

We note that our model cannot take into account potential secondary effects, such as density inhomogeneities or carrier dynamics [e.g., scattering], which we have discussed in the main text. We also surmise that in devices (such as our control sample SB) where the acoustic wave length matches the grain sizes, a single grain might behave as a membrane that undergoes a larger mechanical displacement than that achievable in an uniform graphene sheet.

\end{document}